\documentclass[aps,twocolumn,showpacs,superscriptaddress]{revtex4}
\usepackage{amssymb,graphicx}
\newcommand{\threej}[6]{\left(\matrix{#1 & #2 & #3 \cr     
                                      #4 & #5 & #6 } \right)}

\begin{document}

\title{Classical, semiclassical, and quantum investigations of the 4-sphere
 scattering system}
\author{J\"org Main}
\affiliation{Institut f\"ur Theoretische Physik 1, Universit\"at Stuttgart,
         70550 Stuttgart, Germany}
\author{Erdin\c c At\i lgan}
\author{Howard S. Taylor}
\affiliation{Department of Chemistry, University of Southern California,
         Los Angeles, California 90089}
\author{G\"unter Wunner}
\affiliation{Institut f\"ur Theoretische Physik 1, Universit\"at Stuttgart,
         70550 Stuttgart, Germany}
\date{\today}

\begin{abstract}
A genuinely three-dimensional system, viz.\ the hyperbolic 4-sphere scattering
system, is investigated with classical, semiclassical, and quantum mechanical
methods at various center-to-center separations of the spheres.
The efficiency and scaling properties of the computations are discussed 
by comparisons to the two-dimensional 3-disk system.
While in systems with few degrees of freedom modern quantum calculations are,
in general, numerically more efficient than semiclassical methods, this 
situation can be reversed with increasing dimension of the problem.
For the 4-sphere system with large separations between the spheres, we
demonstrate the superiority of semiclassical versus quantum calculations, 
i.e., semiclassical resonances can easily be obtained even in energy regions 
which are unattainable with the currently available quantum techniques.
The 4-sphere system with touching spheres is a challenging problem for both
quantum and semiclassical techniques.
Here, semiclassical resonances are obtained via harmonic inversion of a
cross-correlated periodic orbit signal.
\end{abstract}


\maketitle

\section{Introduction}
\label{intro:sec}
The breakthrough for the semiclassical quantization of chaotic systems 
was the development of periodic orbit theory \cite{Gut71,Gut90}.
In Gutzwiller's trace formula the density of states is expressed as an
infinite sum over all isolated periodic orbits of the classical system.
Although the periodic orbit theory is in principle valid for systems with 
an arbitrary number of degrees of freedom, applications have, for practical
reasons, so far mostly been restricted to two-dimensional systems.
The main difficulties are, firstly, the numerical periodic orbit search, 
which becomes more difficult in multidimensional systems, and, secondly, 
the fact that the semiclassical trace formula usually does not converge.
The convergence problems can be solved, e.g., with cycle-expansion
\cite{Cvi89,Eck93,Eck95} or harmonic inversion \cite{Mai97c,Mai99c,Mai00}
techniques, and both methods have been successfully applied to the 3-disk
billiard as a prototype model of a two-dimensional hyperbolic scattering
system.
Practical applications of periodic orbit theory to three-dimensional systems
are very rare.
For the three-dimensional Sinai billiard extensive quantum computations have
been performed and the quantum spectra have been analyzed in terms of 
classical periodic orbits \cite{Pri95,Pri00}.
However, no semiclassical eigenstates have been calculated from the set of
periodic orbits.
Semiclassical resonances have been obtained for the three-dimensional 
2- and 3-sphere scattering systems \cite{Hen97} but for these systems
all periodic orbits lie in a one- or two-dimensional subspace.

In this paper will investigate the scattering of a particle on four 
equal spheres centered at the corners of a regular tetrahedron.
Classical and semiclassical as well as quantum mechanical methods will be
applied to the 4-sphere system at various center-to-center separations
of the spheres.
The 4-sphere system can be regarded as the simplest extension of the 3-disk 
repellor to three-dimensional space with a set of genuinely three-dimensional
periodic orbits.
Chaotic properties of the 4-sphere system have been verified experimentally
by the observation of fractal structures via optical light-scattering on
the spheres \cite{Swe99,Swe01}.

When solving two- and three-dimensional systems with both quantum and 
semiclassical methods it is interesting to study the scaling properties 
of the quantization methods with the number of degrees of freedom, and to 
compare the efficiency of the various algorithms.
The numerical effort for the quantization of nonintegrable systems usually
increases strongly with the number of degrees of freedom, and therefore
efficient quantization techniques are highly desirable.
A large variety of quantum mechanical and semiclassical methods have been
developed.
The direct solution of Schr\"odinger's equation is possible, e.g., by
time-dependent wave packet expansions or numerical diagonalization of the
Hamiltonian in a complete basis set.
Exact quantum mechanical calculations usually require storage of 
multidimensional wave functions and a computational effort that grows
exponentially with the number of coupled degrees of freedom.
These methods are therefore feasible for systems with relatively few
degrees of freedom.
As an alternative to exact quantum calculations, approximate, e.g.\
semiclassical, methods can be used.
Gutzwiller's trace formula can be applied to systems with an arbitrary 
number of degrees of freedom, however, the number of periodic orbits and 
the numerical effort needed to find them usually increases very rapidly 
with increasing dimension of the phase space.
As a matter of fact, Gutzwiller's periodic orbit theory has been applied 
almost exclusively to systems with two degrees of freedom, viz.\ the 
anisotropic Kepler problem \cite{Gut90,Tan91}, the hydrogen atom in a 
magnetic field \cite{Tan96}, and two-dimensional billiards 
\cite{Cvi89,Eck95,Gas94,Kea94}.
For these systems direct quantum mechanical computations are usually
more powerful and efficient than the semiclassical calculation of spectra
by means of periodic orbit theory.
The 4-sphere system is an example where semiclassical methods turn out to
be superior to direct quantum mechanical computations \cite{Mai02}, i.e., 
semiclassical resonances can easily be obtained even in energy regions 
which are unattainable with the presently known quantum techniques.

The paper is organized as follows.
In Sec.~\ref{class_dyn:sec} we investigate the classical dynamics of the
4-sphere system.
The symbolic code is introduced and its symmetry reduction by means of
the tetrahedra group, $T_d$, is discussed.
The periodic orbits are found in a systematic way by an efficient numerical
periodic orbit search, and the pruning of orbits at small separations of the
spheres is analyzed.
In Sec.~\ref{semiclassics:sec} we introduce the semiclassical techniques for
periodic orbit quantization, viz.\ the cycle-expansion method, the harmonic
inversion method, and the extension of harmonic inversion to cross-correlated
periodic orbit signals.
In Sec.~\ref{qm:sec} we present the method applied for the exact quantum 
mechanical calculation of the resonances.
In Sec.~\ref{results:sec} we show the results for the semiclassical and
quantum resonances at various separations of the spheres.
The results are discussed with special emphasis on the comparison of the 
efficiency of the various methods.
Some concluding remarks are given in Sec.~\ref{conclusion:sec}.

\section{Classical dynamics: The periodic orbits of the 4-sphere system}
\label{class_dyn:sec}
The 4-sphere system is a genuinely three-dimensional billiard where the 
systematic periodic orbit search is a nontrivial task.
In this section we first develop the symbolic dynamics of orbits and 
the symmetry reduction using the tetrahedra group, and then discuss the 
numerical periodic orbit search and the calculation of the periodic orbit
parameters.

\subsection{Symbolic code and symmetry reduction}
\begin{figure}
\begin{center}
\includegraphics[width=0.95\columnwidth]{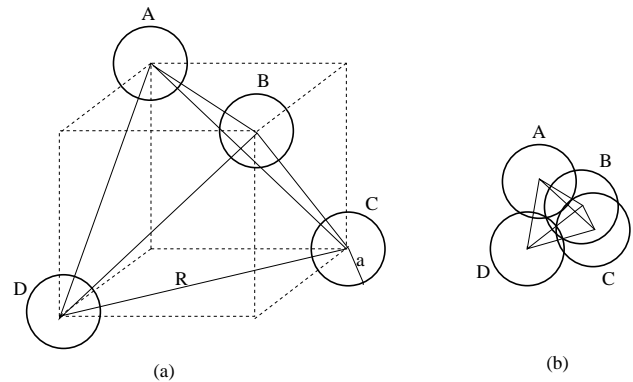}
\end{center}
\caption{The 4-sphere system consists of four equal spheres 
centered at the corners of a regular tetrahedron.
(a) Large center-to-center-separation $R\gg 2a$.
(b) Touching 4-sphere system with $R=2a$.}
\label{fig1}
\end{figure}

The 4-sphere system discussed here consists of four equal spheres with
radius $a$ centered at the corners of a regular tetrahedron.
We choose $a=1$ in what follows.
The system is then solely determined by the center-to-center-separation $R$.
The 4-sphere system with large center-to-center-separation $R\gg 2a$
and with touching spheres ($R=2a$) are shown in Fig.~\ref{fig1} (a) and (b),
respectively.

In full coordinate space each orbit can be described by the infinite 
sequence of spheres where the orbit is scattered.
By labeling the spheres as $\{A, B, C, D\}$, it is possible to code a
periodic orbit as the infinite cycles of a limited length string consisting 
of the sphere labels which we call here the itinerary code of the orbit. 
For a given string length, all combinations of the letters $\{A,B,C,D\}$
correspond to a physical orbit, with the exception that two consecutive 
letters in the itinerary code cannot be identical and, for short 
center-to-center-separation $R\gtrsim 2a$, some orbits may be excluded by 
pruning (see Sec.~\ref{pruning:sec}).
Several itinerary code strings may represent the same periodic orbit or a
similar orbit obtained by a symmetry operation, i.e., rotation or reflection.
For example, the itinerary codes $ABC$ and $BCA$ correspond to the same 
periodic orbit by cyclic permutations, and the orbits $ABC$, $ACD$, $ABD$, 
and $BCD$ can be mapped onto each other by rotations.

By using the symmetry properties $T_d$ of the tetrahedron the system can be 
reduced to its fundamental domain.
The symmetry reduced orbits can be described by a ternary alphabet of symbols
`0', `1', and `2', which are the three fundamental orbits, i.e., the
symmetry reductions of the shortest orbits scattered between two, three, and
four spheres, respectively. 
Therefore, we shall use the symbol `0' for returning back to 
the previous sphere after one reflection, symbol `1' for the reflection to
the other third sphere in the same reflection plane of the orbit, and symbol 
`2' for the reflection to the other forth sphere out of the reflection plane 
of the orbit. 
The reflection plane is defined by the centers of the first three 
{\em different} spheres toward back in the history of the itinerary code 
of the orbit.
The primitive periodic orbits of cycle length $n_p$ in the fundamental domain 
are now given by those periodic sequences of $n_p$ symbols `0', `1', and `2' 
which are free of subcycles (e.g.\ the code $\overline{0101}$ with subcycle 
$\overline{01}$ is not primitive, we will neglect the line indicating 
periodicity in the following).
The periodic orbits do not change by cyclic permutations of the code.
We will choose the code word with the lowest numerical value as the
representative (e.g.\ 0112 instead of 1120).
With these rules every symmetry reduced periodic orbit of the 4-sphere
system is uniquely described by a symbolic code.
However, at small separations of the spheres some physical orbits are pruned
as discussed below in Sec.~\ref{pruning:sec}.

From the $\{0,1,2\}$ code of the symmetry reduced orbits the $\{A,B,C,D\}$
itinerary code can be obtained as follows.
We choose the plane spanned by the spheres $(A,B,C)$ as the initial reflection
plane and start the journey with the sequence $AB$.
Then the rules given above are applied for the symbols 0, 1, and 2 to
guide the orbit to the subsequent spheres.
Note that symbolic codes which contain only the symbols `0' and `1' lie 
in the two-dimensional $(A,B,C)$-plane, i.e., they correspond to the set of 
orbits with a binary symbolic code, which has been well-established for 
the 3-disk \cite{Cvi89,Eck95} and 3-sphere system \cite{Hen97}. 
Orbits including the symbol `2' are genuinely three-dimensional orbits. 
In Table \ref{table1} we present the symbolic codes of all periodic orbits
up to cycle lengths $n_p=3$ of the symmetry reduced code.
\begin{table}
\caption{\label{table1}
Symbolic code $p$ of the symmetry reduced periodic orbits with cycle lengths
$n_p\le 3$ and the itinerary codes $\tilde p$ of the orbits in full coordinate
space.  The column $h_{\tilde p}$ gives the symmetry type of the orbits.}
\begin{center}
\begin{tabular}[t]{l|l|l}
 $p$  &  $\tilde p$  &  $h_{\tilde p}$ \\
\hline
 0   & $AB$           & $\sigma_d$, $C_2$\\
 1   & $ABC$          & $C_3$      \\
 2   & $ABDC$         & $S_4$      \\
 01  & $ABAC$         & $\sigma_d$ \\
 02  & $ABADAC$       & $C_3$      \\
 12  & $ABCDBADC$     & $S_4$      \\
 001 & $ABABCBCAC$    & $C_3$      \\
 002 & $ABABDBDCDCAC$ & $S_4$      \\
 011 & $ABACBC$       & $\sigma_d$ \\
 012 & $ABACDC$       & $C_2$      \\
 021 & $ABADBDCBC$    & $C_3$      \\
 022 & $ABADCDBABCDC$ & $S_4$      \\
 112 & $ABCADC$       & $\sigma_d$ \\
 122 & $ABCDACBDC$    & $C_3$      \\
\end{tabular}
\end{center}
\end{table}
Note that no subcycles and cyclic permutations exist on the list. 
In the second column, the corresponding itinerary codes $\tilde p$ of the 
symbolic codes of column 1 are given, which have been obtained by following
the rules explained above.
The last column in Table~\ref{table1} shows the symmetry classes of the
orbits.
The $T_d$ group has 1 $e$, 3 $C_2$, 8 $C_3$, 6 $S_4$, and 6 $\sigma_d$, 
in total 24 different symmetry elements.
Each orbit (except the one represented by 0) can be assigned by one and only 
one of the symmetry elements $\{e,\sigma_d,C_2,C_3,S_4\}$ of the group $T_d$. 
Note that periodic orbits in the fundamental domain, and thus their symmetry 
reduced symbolic codes, are two-, three-, or four-times shorter than the 
orbits (and the itinerary codes) in the full coordinate space when they belong
to the symmetry class $\{\sigma_d,C_2\}$, $C_3$, or $S_4$, respectively.
The symbolic length of orbits belonging to symmetry class $e$, i.e., the
identity is unchanged under symmetry reduction.

\subsection{Numerical periodic orbit search}
Each trajectory of the 4-sphere system is completely determined by the
reflection points on the surfaces of the spheres, which on each sphere
can be described by two spherical coordinates $\theta$ and $\phi$.
For a given itinerary code arbitrarily chosen reflection points on the 
spheres connected by straight lines in the correct order result in a
periodic but not necessarily a physical orbit.
The true physical orbit, for which the incident and reflection angle at 
each reflection point must coincide, can be obtained by direct application 
of Hamilton's principle, i.e., the orbital length, which is proportional to
the classical action, becomes a minimum when the reflection points are varied.
The length function of an orbit with a total number of $N$ reflection points
depends on the $2N$ variables $\{\theta_i, \phi_i\}$ with $i=1,\dots,N$.
Numerically, the minimizing of the length 
\begin{equation}
 L = L(\theta_1, \phi_1, \theta_2, \phi_2, \dots, \theta_N, \phi_N)
\end{equation}
can be achieved by applying the well established quasi-Newton method 
\cite{NumRec}, which is implemented, e.g., in the NAG-library \cite{NAG}.
The required gradient of the length function, $\nabla L$, has been derived
analytically.

As mentioned above the length of periodic orbits in full coordinate space
can be two, three, or four times the length of the corresponding symmetry
reduced orbit in the fundamental domain (see Table~\ref{table1}).
As the required computational effort for the quasi-Newton method increases
rapidly with the dimensionality of the problem, it is desirable to exploit
the symmetry properties of the tetrahedra group and to directly search
for the periodic orbits in the fundamental domain.
To this end for a symmetry reduced orbit with cycle length $n_p$ the
reflection point on the sphere $n_p+1$ is associated with the reflection
point on the first sphere by an appropriate transformation, i.e., one of
the 24 possible symmetry transformations of the tetrahedra group, $T_d$.
The length minimization is now applied to the trajectory segments between
the first sphere and sphere $n_p+1$, i.e., the dimensionality of the
length minimization of periodic orbits in the fundamental domain is reduced
to $2n_p$ for all primitive orbits with cycle length $n_p$.

Once a periodic orbit has been found its orbital parameters required
for semiclassical periodic orbit quantization can be calculated.
The most important ones are the monodromy matrix and the Maslov index of
the orbit.
The Maslov index increases by 2 at each reflection on a hard sphere, i.e.,
$\mu_{\rm po}=2n_p$ for an orbit with cycle length $n_p$.
The calculation of the monodromy matrix ${\mathbf M}_{\rm po}$ for the 
periodic orbits of three-dimensional billiards has been investigated in 
Refs.~\cite{Pri00,Sie98}.
${\mathbf M}_{\rm po}$ is a symplectic $(4\times 4)$ matrix with eigenvalues
$\lambda_1$, $1/\lambda_1$, $\lambda_2$, and $1/\lambda_2$.
For the hyperbolic 4-sphere system $\lambda_1$ and $\lambda_2$ are either
both real or the orbits are loxodromic, i.e., the eigenvalues of
${\mathbf M}_{\rm po}$ are a quadruple 
$\{\lambda, 1/\lambda, \lambda^\ast, 1/\lambda^\ast\}$ with $\lambda$ being
a complex number.
For the 4-sphere system with radius $a=1$ and center-to-center separation
$R=6$ the orbital lengths and stability parameters for all primitive periodic
orbits with cycle length $n_p\le 3$ are presented in Table~\ref{table2}.
\begin{table}
\caption{\label{table2}
Parameters of the symmetry reduced primitive periodic orbits $p$ with 
cycle length $n_p\le 3$ of the 4-sphere system with radius $a=1$ and 
center-to-center separation $R=6$.}
\begin{center}
\begin{tabular}[t]{lcrrrrr}
  \multicolumn{1}{l}{$p$} &
  \multicolumn{1}{c}{$h_p$} &
  \multicolumn{1}{c}{$L$} &
  \multicolumn{1}{c}{Re~$\lambda_1$} &
  \multicolumn{1}{c}{Im~$\lambda_1$} &
  \multicolumn{1}{c}{Re~$\lambda_2$} &
  \multicolumn{1}{c}{Im~$\lambda_2$} \\
\hline
 0 &$\sigma_d, C_2$ &4.000000 &  9.89898 &  0.00000 &  9.89898 &  0.00000\\
 1   & $C_3$      &  4.267949 & -11.7715 &  0.00000 &  9.28460 &  0.00000\\
 2   & $S_4$      &  4.296322 & -4.52562 &  9.49950 & -4.52562 & -9.49950\\
 01  & $\sigma_d$ &  8.316529 & -124.095 &  0.00000 &  88.4166 &  0.00000\\
 02  & $C_3$      &  8.320300 & -37.1479 &  98.0419 & -37.1479 & -98.0419\\
 12  & $S_4$      &  8.567170 &  117.644 &  0.00000 & -102.992 &  0.00000\\
 001 & $C_3$      & 12.321747 & -1240.54 &  0.00000 &  868.915 &  0.00000\\
 002 & $S_4$      & 12.322138 & -353.853 &  976.176 & -353.853 & -976.176\\
 011 & $\sigma_d$ & 12.580808 &  1449.55 &  0.00000 &  824.981 &  0.00000\\
 012 & $C_2$      & 12.617350 &  1192.83 &  0.00000 & -1020.66 &  0.00000\\
 021 & $C_3$      & 12.584068 &  1201.43 &  0.00000 & -996.800 &  0.00000\\
 022 & $S_4$      & 12.619948 & -755.582 &  804.976 & -755.582 & -804.976\\
 112 & $\sigma_d$ & 12.835715 & -496.339 &  1038.46 & -496.339 & -1038.46\\
 122 & $C_3$      & 12.863793 & -1100.56 &  0.00000 &  1219.28 &  0.00000
\end{tabular}
\end{center}
\end{table}
For that sphere separation ($R=6a$) we have calculated the complete set of 
primitive periodic orbits with symbol lengths $n_p\le 14$, numbering 533830 
orbits in total.
For sphere separation $R=2.5\, a$ we also calculated all primitive periodic 
orbits with symbol lengths $n_p\le 14$, and in addition all orbits with
symbol lengths $n_p\le 22$ and physical lengths $L\le 12$, which allows for
the construction of a periodic orbit signal with length $L_{\rm max}=12$
used for the semiclassical quantization in Sec.~\ref{d2.5:sec}.

\subsection{Pruning of orbits}
\label{pruning:sec}
\begin{figure}
\begin{center}
\includegraphics[width=0.8\columnwidth]{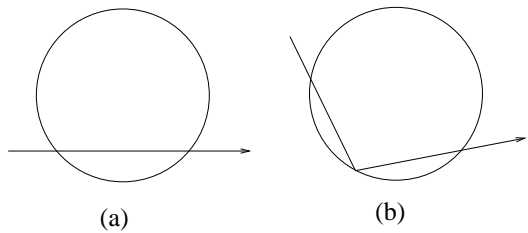}
\end{center}
\caption{Sketch of the two types of pruning occurring in the 4-sphere
system at small separation between the spheres: (a) An orbital segment
passes through one of the spheres. (b) A reflection occurs inside a sphere.}
\label{fig2}
\end{figure}

For center-to-center separations $R>2.0482\, a$ between the spheres there
is a one to one correspondence between the symbolic codes and the primitive
periodic orbits.
However, when the separation is reduced below that value some orbits become 
unphysical, i.e., the symbolic dynamics is pruned.
The pruning of orbits has been investigated in detail for the 3-disk
scattering system \cite{Han92,Han93}.
For the 4-sphere system the mechanism is similar:
As illustrated in Fig.~\ref{fig2}, an orbital segment may (a) pass 
through one of the spheres, or (b) a reflection may occur inside one
of the spheres.
For a periodic orbit search at small separation between the spheres
all orbits obtained numerically by minimizing the length must be
checked whether pruning occurs or not.
For touching spheres ($R=2a$) all pruned orbits with symbol lengths $n_p\le 7$
and their pruning types (a) or (b) are presented in Table~\ref{table3}.
\begin{table}
\caption{All pruned orbits with cycle length $n_p\le 7$ and their pruning 
types (a) or (b) (see Fig.~\ref{fig2}) of the 4-sphere system 
with touching spheres, $R=2a$.}
\begin{center}
\begin{tabular}{l|c}
symbolic code & pruning type \\
\hline 
00021   & a \\
000011  & a \\
000021  & a \\
000002  & b \\
0000001 & b \\
0000011 & a \\
0000021 & a \\
0000002 & b \\
\end{tabular}
\end{center}
\label{table3}
\end{table}
Pruning exists for orbits with symbol lengths $n_p\ge5$, i.e.,
the symbolic dynamics is complete only for $n_p\le 4$.
Furthermore, periodic orbits with long heads of `0' symbols in the code 
can have accumulation points at finite values of the physical length $L$.
It is impossible to find all orbits beyond the first accumulation point.
We have searched for all periodic orbits of the touching 4-sphere system
with physical lengths $L\le L_{\rm max}=3.6$, symbol lengths $n_p\le 60$, 
and with the total number of `1' and `2' symbols in the symbolic code 
restricted to $n_1+n_2\le 10$, resulting in about 2.8 million primitive 
periodic orbits.

The semiclassical quantization by harmonic inversion of a cross-correlated
periodic orbit signal (see Sec.~\ref{cross-corr:sec}) requires the
knowledge of the expectation values 
of various linearly independent classical observables $A$ along the 
periodic orbits \cite{Mai99a,Mai99b}.
We have chosen the observables $A_1=r^2$ and $A_2=L^2$, i.e., we have 
calculated the averaged squared distance and squared angular momentum of 
the periodic orbits of the touching 4-sphere system.

\section{Semiclassical periodic orbit theory}
\label{semiclassics:sec}
We now wish to calculate semiclassically the resonances of the 4-sphere
scattering system by application of periodic orbit theory.
Gutzwiller's trace formula \cite{Gut90} expresses the quantum mechanical
response function
\begin{equation}
 g_{\rm qm}(E) = \sum_n \frac{1}{E-E_n+{\rm i}\epsilon}
\label{g_qm}
\end{equation}
in terms of the periodic orbits of the underlying classical system, i.e.,
\begin{equation}
 g_{\rm scl}(E) = g_0(E) + \sum_{\rm po} A_{\rm po}(E) \,
  {\rm e}^{{\rm i}S_{\rm po}(E)/\hbar} \; ,
\label{g_scl}
\end{equation}
where $g_0(E)$ is a smooth function of the energy and $A_{\rm po}(E)$ and
$S_{\rm po}(E)$ are the periodic orbit amplitudes (including phase information
given by the Maslov indices) and classical actions, respectively.
For billiards the classical action depends linearly on the length of
the trajectory and the wave number $k=\sqrt{2ME}/\hbar$ with $M$ being the
particle mass.
For the three-dimensional 4-sphere system the periodic orbit sum as
a function of the wave number $k$ reads
\begin{equation}
 g(k) = \sum_p \sum_{r=1}^\infty
   \frac{w_p(-1)^{rn_p}L_p{\rm e}^{{\rm i}krL_p}}
        {\sqrt{|(2-\lambda_{p,1}^r-\lambda_{p,1}^{-r})
                (2-\lambda_{p,2}^r-\lambda_{p,2}^{-r})|}} \; ,
\label{g_k}
\end{equation}
where $n_p$ is the cycle length, $L_p$ the physical length, $\lambda_{p,i}$ 
are the eigenvalues of the monodromy matrix, and $r$ is the repetition number
of the primitive periodic orbit $p$. 
The weight factors $w_p$ result from the symmetry decomposition of the
system \cite{Cvi93} and depend on the chosen irreducible subspace of the 
spectrum and the symmetry of the periodic orbits.
For the tetrahedra group, $T_d$, the values of the weight factors $w_p$
are given in Table \ref{table4}.
\begin{table}
\caption{Weight factors $w_p$ for the periodic orbit sum (Eq.~(\ref{g_k}))
of the 4-sphere system with symmetries of the tetrahedra group, $T_d$.}
\begin{center}
\begin{tabular}[t]{l|rrrrr}
 $T_d$ & $~e$   & $C_3$ & $C_2$ & $S_4$ & $\sigma_d$ \\
\hline
 $A_1$ &   1   &   1   &   1   &   1   &   1  \\
 $A_2$ &   1   &   1   &   1   &  -1   &  -1  \\
 $E$   &   2   &  -1   &   2   &   0   &   0  \\
 $T_1$ &   3   &   0   &  -1   &   1   &  -1  \\
 $T_2$ &   3   &   0   &  -1   &  -1   &   1
\end{tabular}
\end{center}
\label{table4}
\end{table}
In the following we will concentrate on the subspace $A_1$, where the weight
factors of all orbits are $w_p=1$.

The semiclassical resonances of the 4-sphere system are given by the poles 
of the function $g(k)$.
However, it is well known that the periodic orbit sum (\ref{g_k}) does not 
converge in those regions where the physical poles are located, and special
techniques must be applied to obtain an analytical continuation of the
periodic orbit sum (\ref{g_k}).
For the 3-disk system with large center-to-center separation $R=6a$
the cycle-expansion method \cite{Cvi89,Eck95,Wir99} and harmonic inversion 
techniques \cite{Mai97c,Mai99c} have proven to be powerful approaches for 
overcoming the convergence problems of the periodic orbit sum, and both 
methods can also be successfully applied to the 4-sphere system.
However, when pruning of orbits sets in at small separations, and in 
particular in the case of touching disks or spheres, the situation is 
much more difficult and subtle, since the direct application of the 
cycle-expansion method fails.
The two-dimensional closed 3-disk billiard is a bound system, where a few 
semiclassical eigenenergies have been obtained in Ref.~\cite{Tan91} using
the cycle-expansion in combination with a functional equation.
This method is not valid for open systems and cannot be extended to the
4-sphere system which remains open even in the case of touching spheres
\cite{Swe99,Swe01}.
Nevertheless, the harmonic inversion of cross-correlated periodic orbit 
signals \cite{Mai99a,Mai99b} has been successfully applied to the closed 
3-disk system \cite{Wei02a,Wei02b} and this method will also serve
as a powerful tool for the three-dimensional 4-sphere system.
We will now introduce the quantization methods.
Applications to the 4-sphere system and comparisons with quantum mechanical
results will be presented in Sec.~\ref{results:sec}.

\subsection{The cycle-expansion method}
\label{cycle-exp:sec}
The periodic orbit sum in Gutzwiller's trace formula does usually not
converge in the energy regions of physical interest.
However, for some systems, e.g., the 3-disk scattering billiard,
semiclassical energies or resonances can be obtained with the help
of the cycle-expansion method \cite{Cvi89,Eck93,Eck95}.
If the periodic orbits can be associated to a symbolic dynamics
the Gutzwiller-Voros zeta function \cite{Gut90,Vor88} can be
expanded according to increasing cycle length of the orbits.
In this expansion the contributions of long periodic orbits may be
approximately shadowed by the combined contributions of shorter orbits.
In this case the cycle-expansion can converge rapidly.

For billiards the Gutzwiller-Voros zeta function can be written as
\begin{equation}
 Z_{\rm GV}(k;z) = \exp\left\{-\sum_p\sum_{r=1}^\infty\frac{1}{r}
  \frac{(-z)^{rn_p}{\rm e}^{{\rm i}rkL_p}}
       {\sqrt{|\det({\mathbf M}_p^r-{\mathbf 1})|}}\right\} \; ,
\label{Z_GV}
\end{equation}
with an additional parameter $z$ which must be set to $z=1$.
The cycle-expansion is achieved by taking $z$ as a book-keeping variable
and expanding Eq.~(\ref{Z_GV}) as a truncated power series in $z$.
The semiclassical resonances are obtained as the zeros (in the variable $k$)
of the cycle-expanded zeta function (\ref{Z_GV}) with again $z=1$.
In our computations for the 4-sphere system we use cycle-expansions up
to order $n_{\rm max}=12$.

\subsection{Semiclassical quantization by harmonic inversion}
\label{hi:sec}
An alternative method for semiclassical quantization is based on the
observation that the extraction of eigenvalues from Gutzwiller's trace formula
can be reformulated as a signal processing task \cite{Mai97c,Mai99c,Mai00}.
The harmonic inversion method is briefly explained as follows.
The Fourier transform of the function $g(k)$ in Eq.~(\ref{g_k}) yields
the semiclassical signal
\begin{equation}
 C^{\rm sc}(L) = \sum_p \sum_{r=1}^\infty
   \frac{(-1)^{rn_p} L_p}
        {\sqrt{|\det({\mathbf M}_p^r-{\mathbf 1})|}} \delta(L-rL_p) \; ,
\label{C_sc}
\end{equation}
as a sum of $\delta$ functions.
The central idea of semiclassical quantization by harmonic inversion is
to adjust the semiclassical signal $C^{\rm sc}(L)$ with finite length
$L\le L_{\rm max}$ to its quantum mechanical analogue
\begin{eqnarray}
 C^{\rm qm}(L)
 &=& \frac{\rm i}{2\pi} \int_{-\infty}^{+\infty} \sum_n
   \frac{d_n}{k-k_n+{\rm i}\epsilon} {\rm e}^{-{\rm i}kL} {\rm d}k \nonumber \\
 &=& \sum_n d_n {\rm e}^{-{\rm i}k_nL} \; ,
\label{C_qm}
\end{eqnarray}
where the amplitudes $d_n$ and the semiclassical eigenvalues $k_n$ 
are free adjustable complex parameters.
This is achieved by signal processing \cite{Wal95,Man97} of the
semiclassical signal $C^{\rm sc}(L)$.
Numerical recipes for extracting the parameters $\{d_n,k_n\}$ by harmonic 
inversion of the $\delta$ function signal (\ref{C_sc}) are given 
in \cite{Mai00,Bar01}.

\subsection{Harmonic inversion of cross-correlated periodic orbit signals}
\label{cross-corr:sec}
The method of semiclassical quantization by harmonic inversion of 
cross-correlated periodic orbit signals is a generalization of the
quantization scheme presented in Sec.~\ref{hi:sec}.
The idea is to use the classical average values of a set of linearly
independent classical observables to construct a cross-correlated signal,
whose informational content is significantly increased as compared to the 
one-dimensional signal, and therefore should lead to semiclassical spectra 
with improved resolution.

The numerical tools for the harmonic inversion of cross-correlated periodic
orbit signals have already been well established \cite{Mai99b}, and therefore 
we only briefly review the basic ideas and refer the reader to the literature
for details.
For simplicity but without loss of generality, we focus on billiard 
systems, where the shape of the orbits is independent of the energy $E$,
and the classical action of orbits reads $S=\hbar kL$, with $k$ the wave
number and $L$ the physical length.
The starting point is to introduce a weighted response function
in terms of $k$
\begin{equation}
   g_{\alpha\alpha'}(k)
 = \sum_n {b_{\alpha n}b_{\alpha' n} \over k-k_n+{\rm i}\epsilon}\ ,
\label{g_ab_qm}
\end{equation}
where $k_n$ is the eigenvalue of the wave number of eigenstate $|n\rangle$
and
\begin{equation}
 b_{\alpha n} = \langle n|\hat A_\alpha|n\rangle
\end{equation}
are the diagonal matrix elements of a chosen set of $N$ linearly independent
operators $\hat A_\alpha$, $\alpha=1,2,\dots, N$.
The Fourier transform of (\ref{g_ab_qm}) yields the $N\times N$
cross-correlated signal
\begin{eqnarray}
 C_{\alpha\alpha'}(L)
 &=& \frac{\rm i}{2\pi}\int_{-\infty}^{+\infty}
      g_{\alpha\alpha'}(k){\rm e}^{-{\rm i}kL}{\rm d}k  \nonumber \\
 &=& \sum_n b_{\alpha n}b_{\alpha' n} {\rm e}^{-{\rm i}k_nL} \; .
\label{C_ab_qm}
\end{eqnarray}
A semiclassical approximation to the cross-correlated signal (\ref{C_ab_qm})
has been derived in \cite{Mai99a,Hor00}.
The cross-correlated periodic orbit signal reads
\begin{equation}
   C_{\alpha\alpha'}^{\rm sc}(L)
 = \sum_p \sum_{r=1}^\infty
   \frac{a_{\alpha,p}\, a_{\alpha',p} (-1)^{rn_p} L_p}
        {\sqrt{|\det({\mathbf M}_p^r-{\mathbf 1})|}} \delta(L-rL_p) \; ,
\label{C_ab_sc}
\end{equation}
where $r$ is the repetition number counting the traversals of the primitive 
orbit, and ${\mathbf M}_p$ is the monodromy matrix of the primitive periodic
orbit.
The weight factors $a_{\alpha,p}$ are classical averages over the 
periodic orbits
\begin{equation}
 a_{\alpha,p} = {1\over L_p} \int_0^{L_p}
  A_\alpha({\bf q}(L),{\bf p}(L)) {\rm d}L \; ,
\label{a_po}
\end{equation}
with $A_\alpha({\bf q},{\bf p})$ the Wigner transform of the operator
$\hat A_\alpha$.
Semiclassical approximations to the eigenvalues $k_n$ and eventually also
to the diagonal matrix elements $\langle n|\hat A_\alpha|n\rangle$ are
now obtained by adjusting the semiclassical cross-correlated periodic orbit
signal (\ref{C_ab_sc}) to the functional form of the quantum signal
(\ref{C_ab_qm}).
The numerical tool for this procedure is an extension of the harmonic 
inversion method to the signal processing of cross-correlation functions 
\cite{Nar97,Man98}.
The advantage of using the cross-correlation approach is based on the 
realization that the total amount of independent information contained in
the $N\times N$ signal is $N(N+1)$ multiplied by the length of the signal, 
while the total number of unknowns (here $b_{\alpha n}$ and $k_n$) is $(N+1)$
times the total number of poles $k_n$. 
Therefore the informational content of the $N\times N$ signal per unknown 
parameter is increased (as compared to the one-dimensional signal) by 
roughly a factor of $N$, and the cross-correlation approach should lead to 
a significant improvement of the resolution.

\section{Quantum calculations}
\label{qm:sec}
Schr\"odinger's equation for the 3-disk or the 4-sphere system is a free wave
equation in two or three dimensions, $[\Delta+k^2]\Psi({\bf k})=0$, with 
Dirichlet boundary conditions, i.e., $\Psi({\bf k})=0$ on the surface of the 
disks or spheres, respectively.
Although the problem looks simple the solution is a nontrivial task and, 
most importantly, the numerical effort increases extremely rapidly with
the dimension of the system.

For the 3-disk system the exact quantum resonances can be obtained as
roots of the equation \cite{Gas89,Wir99}
\begin{equation}
 \det {\mathbf M}(k)_{mm'}^{\textrm{\scriptsize 3-disk}} = 0 \; ,
\label{M_3disk}
\end{equation}
with $m$ and $m'$ nonzero integer numbers which can be truncated by an
upper angular momentum $m_{\rm max} \gtrsim 1.5\, ka$ \cite{Wir99}.
With matrices ${\mathbf M}(k)_{mm'}^{\textrm{\scriptsize 3-disk}}$ of 
dimension up to $\sim (400\times 400)$, Eq.~(\ref{M_3disk}) allows for 
the efficient numerical calculation of resonances in the region 
$0\le {\rm Re}\, ka \le 250$.
The matrix elements of ${\mathbf M}(k)_{mm'}^{\textrm{\scriptsize 3-disk}}$
in Eq.~(\ref{M_3disk}) can be written analytically in terms of Bessel and
Hankel functions.
Explicit expressions are given in \cite{Wir99}.
The quantum resonances are obtained by a numerical root search in the 
complex $k$-plane \cite{NumRec}.

Similarly, exact quantum resonances of the three-dimensional 4-sphere 
scattering system can be obtained as roots of the equation
\begin{equation}
 \det {\mathbf M}(k)_{lm,l'm'}^{\textrm{\scriptsize 4-sphere}} = 0 \; ,
\label{M_4sphere}
\end{equation}
with $0\le l,l'\le l_{\rm max}$ and $m,m'=0,3,6,9,\dots,l_{\rm max}$
for the subspace $A_1$ and $A_2$.
An explicit expression for the matrix elements of 
${\mathbf M}(k)_{lm,l'm'}^{\hbox{\scriptsize\rm 4-sphere}}$
has been derived \cite{Hen97} and reads
\begin{eqnarray}
     {\mathbf M}(k)_{lm,l'm'}^{\hbox{\scriptsize\rm 4-sphere}}
 &=& \delta_{ll'} \delta_{mm'} + \frac{3}{2}\sqrt{4\pi}{\rm i}^{l'-l}
     \frac{j_l(ka)}{h_{l'}^{(1)}(ka)} g_m g_{m'} \nonumber \\
 &\times&
 \sum_{\tilde l=0}^\infty C(l,m,l',m',\tilde l;\theta_0,\beta_0)
 h^{(1)}_{\tilde{l}}(kR)
\label{4s-M}
\end{eqnarray}
with
\begin{eqnarray}
 & & C(l,m,l',m',\tilde l;\theta_0,\beta_0)  \nonumber \\
 &=& \sum_{M=-l'}^{l'} {\rm i}^{\tilde l}
       \sqrt{(2l+1)(2l'+1)(2\tilde{l}+1)}
      \threej{\tilde{l}}{l'}{l}{0}{0}{0}  \;  \nonumber \\
 &\times&  (-1)^M \left( d^{l'}_{m'M}(\beta_0) \pm
           (-1)^{m'} d^{l'}_{-m',M}(\beta_0)
            \right) \;  \nonumber \\
 &\times&  \bigg[ (-1)^m Y_{\tilde{l},m-M}(\theta_0,0)
       \threej{\tilde{l}}{l'}{l}{m-M}{M}{-m}  \nonumber \\
 & & \pm  Y_{\tilde{l},-m-M}(\theta_0,0)
       \threej{\tilde{l}}{l'}{l}{-m-M}{M}{m} \bigg] \; ,
\label{C_def}
\end{eqnarray}
where the $\pm$-signs refer to the subspace $A_1$ and $A_2$, respectively.
The angles $\theta_0$ and $\beta_0$ in (\ref{C_def}) are obtained from
\[
\begin{array}{ll}
  \cos(\theta_0) = - \frac{2}{\sqrt{6}} \; , \quad &
  \sin(\theta_0) \; = \; \frac{1}{\sqrt{3}} \; ,  \\[2ex]
  \cos(\beta_0) =  - \frac{1}{3}  \; , &
  \sin(\beta_0) \; = \; \frac{2}{3} \sqrt{2} \; ,
\end{array}
\]
and the $d^j_{mm'}(\beta)$ are the matrix elements of 
finite rotations \cite{Edm57},
\[
  d^j_{mm'}(\beta) = \langle{jm}| {\rm e}^{-{\rm i}\beta J_y}|{jm'}\rangle \; .
\]
The large brackets in (\ref{C_def}) refer to $3j$-symbols \cite{Edm57},
and the values of $g_m$ are defined as
\[
  g_m = \cases{ 1/\sqrt{2} & for $m=0$  \cr
                         1 & for $m=3,6,9,\ldots,l$ \cr
                         0 & otherwise.  }
\]
Note that $g_{m=0}$ should read $g_0=1/\sqrt{2}$ instead of $\sqrt{2}$ in 
Eq.~(38) of \cite{Hen97}.
Similar as for the 3-disk system the angular momentum 
(quantum numbers $l$ and $l'$ in Eq.~(\ref{M_4sphere})) can be truncated at 
$l_{\rm max} \gtrsim 1.5\, ka$ to achieve convergence of the calculation.

It is important to note that the computation of the quantum mechanical 
resonances of the three-dimensional 4-sphere scattering system becomes
much more expensive than for the two-dimensional 3-disk system.
First of all, the calculation of each matrix element
${\mathbf M}(k)_{lm,l'm'}^{\textrm{\scriptsize 4-sphere}}$ in (\ref{4s-M})
requires the summation over quantum numbers $\tilde l$ and (via (\ref{C_def}))
$M$.
To accelerate the calculation of the matrix (\ref{4s-M}) at various values
of $k$ we have calculated and stored the values of 
$C(l,m,l',m',\tilde l;\theta_0,\beta_0)$ in (\ref{C_def}) separately.
Eq.~(\ref{C_def}) does not depend on the sphere separation $R$,
and therefore the stored $C$-values can be used in calculations of 
spectra with arbitrary $R$.
However, the calculation of the matrix elements in (\ref{4s-M}) still
requires the summation over $\tilde l$.

The second problem of solving Eq.~(\ref{M_4sphere}) is the scaling of the 
dimension of the matrix 
${\mathbf M}(k)_{lm,l'm'}^{\textrm{\scriptsize 4-sphere}}$, which is
an $N\times N$ matrix with 
\[
 N=\frac{1}{6}(l_{\rm max}+2)(l_{\rm max}+3) \; ,
\]
i.e., $N$ scales as $N\sim k^2$ for the 4-sphere system, as compared to
$N\sim k$ for the 3-disk system, Eq.~(\ref{M_3disk}).
For example, in the region $ka\approx 200$ the required matrix dimension is 
$N\gtrsim 300$ for the 3-disk system, as compared to $N\gtrsim 15000$ for 
the 4-sphere system.
For the 4-sphere system with center-to-center separations $R=6a$, $R=2.5\, a$,
and the touching spheres $R=2a$ we have computed the quantum resonances
in the region $0\le {\rm Re}\, ka\le 60$ by solving Eq.~(\ref{M_4sphere})
with matrices of dimension up to $(1751\times 1751)$.
The results will be presented in Sec.~\ref{results:sec}.
With currently available computer technology it is impossible to 
significantly extend the quantum calculations for the 4-sphere system 
to the region ${\rm Re}\, ka\gg 60$ using Eq.~(\ref{M_4sphere}).
The efficiency of the semiclassical and quantum methods for the 4-sphere
system will be compared and discussed in Sec.~\ref{efficiency:sec}.

\section{Results and discussion}
\label{results:sec}
We will now present and discuss the results of our semiclassical and quantum
computations for the 4-sphere system with large sphere separation $R=6a$,
intermediate separation $R=2.5\, a$, and touching spheres, $R=2a$.
In Sec.~\ref{efficiency:sec} we will compare and discuss the efficiency
of the various quantization methods.

\subsection{Sphere separation $R=6a$}
\label{d6:sec}
\begin{figure}
\begin{center}
\includegraphics[width=0.95\columnwidth]{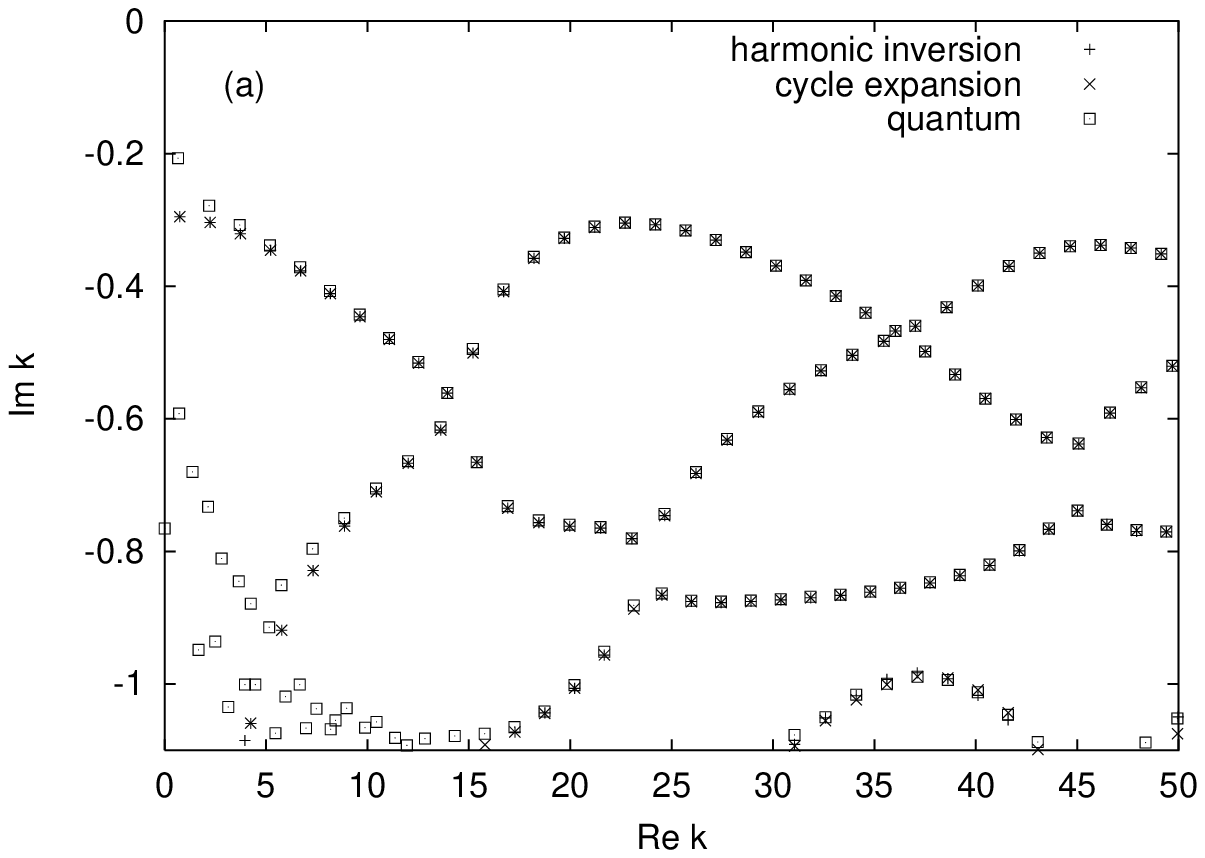}
\includegraphics[width=0.95\columnwidth]{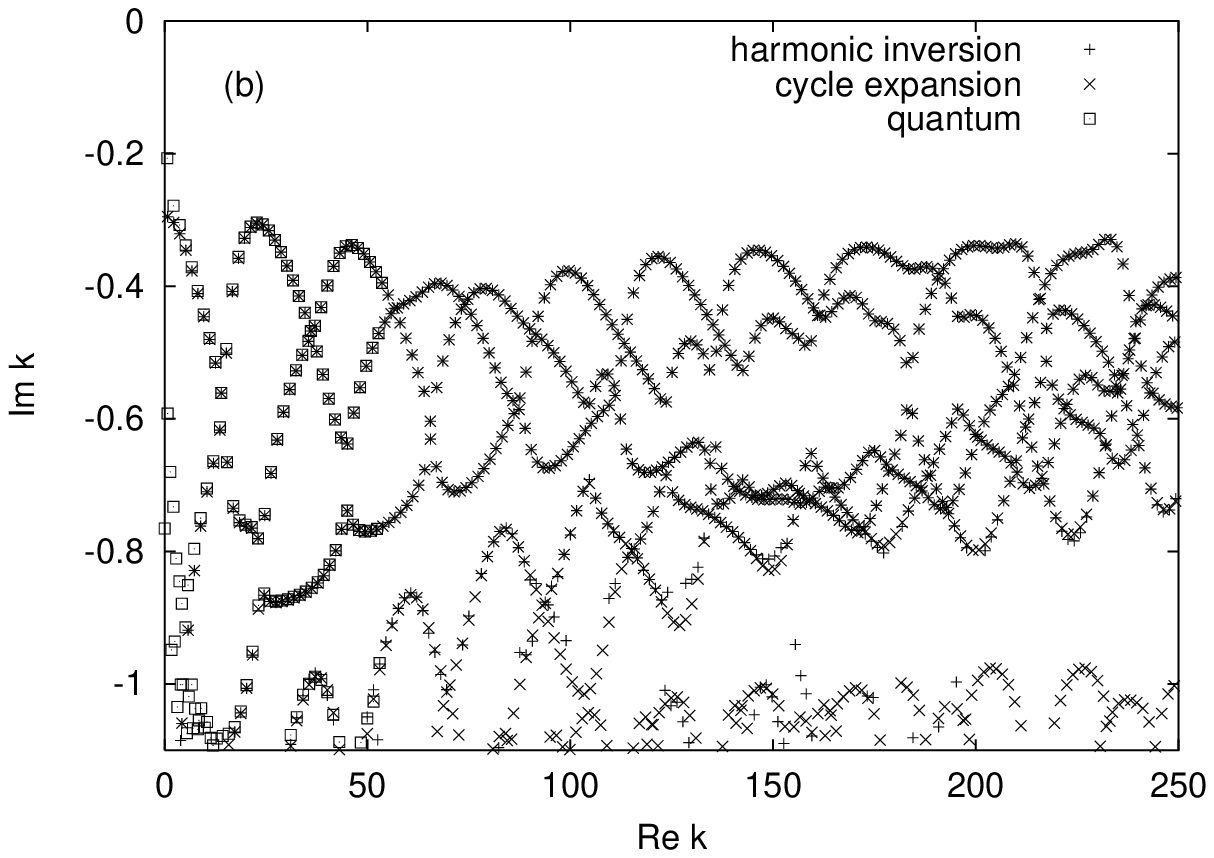}
\end{center}
\caption{$A_1$-resonances in the complex $k$-plane of the 4-sphere system
with radius $a=1$ and center-to-center separation $R=6$. 
Squares: Quantum computations. 
Crosses and plus symbols: Semiclassical resonances obtained by cycle-expansion
and harmonic inversion methods, respectively.}
\label{fig3}
\end{figure}

The quantum mechanical and semiclassical $A_1$-resonances of the 4-sphere
system with radius $a=1$ and center-to-center separation $R=6$ are presented
in Fig.~\ref{fig3}.
The quantum resonances marked by the squares have been obtained by solving
Eq.~(\ref{M_4sphere}) with matrices 
${\mathbf M}(k)_{lm,l'm'}^{\textrm{\scriptsize 4-sphere}}$ of dimension
up to $(1134\times 1134)$, which is sufficient only to obtain converged
results in the region ${\rm Re}~k\lesssim 50$ (see Fig.~\ref{fig3}a).
By contrast, the semiclassical resonances can easily be obtained in a much
larger region, e.g., ${\rm Re}~k\le 250$ shown in Fig.~\ref{fig3}b.
The crosses mark the zeros of the cycle-expanded Gutzwiller-Voros 
zeta function (\ref{Z_GV}).
The cycle-expansion has been truncated at cycle length $n_{\rm max}=7$,
which means that a total set of just 508 primitive periodic orbits are
included in the calculation.
The plus symbols mark the semiclassical resonances obtained by harmonic
inversion of the periodic orbit signal (\ref{C_sc}) with signal length
$L_{\rm max}=60$ constructed from the set of 533830 primitive periodic orbits
with cycle lengths $n_p\le 14$.

In the region ${\rm Re}~k\le 50$ (Fig.~\ref{fig3}a) the quantum and
semiclassical resonances agree very well, with a few exceptions.
The first few quantum resonances in the uppermost resonance band are narrower,
i.e., closer to the real axis than the corresponding semiclassical resonances.
A similar discrepancy between quantum and semiclassical resonances has already
been observed in the 3-disk system \cite{Eck95,Wir99}.
Furthermore, in the region ${\rm Re}~k<15$ and ${\rm Im}~k<-0.5$ several
quantum resonances have been found (see the squares in Fig.~\ref{fig3}a),
which seem not to have any semiclassical analogue.
These resonances are related to the diffraction of waves at the spheres, and
its semiclassical description requires an extension of Gutzwiller's trace
formula and the inclusion of diffractive periodic orbits \cite{Vat94,Ros96}.
The semiclassical resonances obtained by either harmonic inversion or the
cycle-expansion method (the plus symbols and crosses in Fig.~\ref{fig3}b,
respectively) are generally in perfect agreement, except for the very 
broad resonances that lie deep in the complex plane, i.e., in the 
region ${\rm Im}~k\lesssim -0.8$.

\subsection{Sphere separation $R=2.5\, a$}
\label{d2.5:sec}
The semiclassical quantization becomes more and more demanding with
decreasing separation between the spheres.
The reason is that the shadowing of longer orbits by combinations of
shorter orbits in the cycle-expanded Gutzwiller-Voros zeta function
becomes less accurate and the construction of the periodic orbit signal 
of length $L\le L_{\rm max}$ used for the harmonic inversion method
requires more and more periodic orbit data.
However, both semiclassical quantization techniques, i.e., cycle-expansion
and harmonic inversion can still be successfully applied at significantly
reduced separation between the spheres.

\begin{figure}
\begin{center}
\includegraphics[width=0.95\columnwidth]{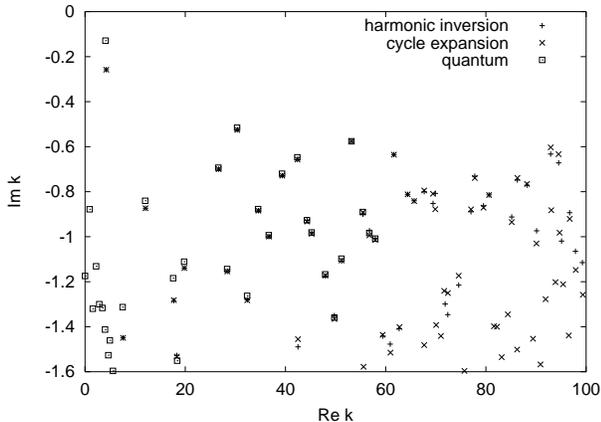}
\end{center}
\caption{$A_1$-resonances in the complex $k$-plane of the 4-sphere system
with radius $a=1$ and center-to-center separation $R=2.5$.  Squares: Quantum
computations.  Crosses and plus symbols: Semiclassical resonances obtained 
by cycle-expansion and harmonic inversion methods, respectively.}
\label{fig4}
\end{figure}

As an example of an intermediate sphere separation we discuss the case
$R=2.5\, a$, where the spheres are rather close, however, the symbolic dynamics
of the periodic orbits is still complete, i.e., no orbits are pruned.
The graphical comparison of the quantum mechanical and semiclassical 
resonances in the region $0\le {\rm Re}~ka\le 100$ is given 
in Fig.~\ref{fig4}.
The semiclassical resonances shown as plus symbols have been obtained by
harmonic inversion of a periodic orbit signal of length $L_{\rm max}=12$.
The signal has been constructed using all primitive periodic orbits with
symbol lengths $n_p\le 14$ and parts of the orbits with symbol lengths
$15\le n_p\le 22$, in total a set of about 4.6 million orbits.
The crosses in Fig.~\ref{fig4} mark the semiclassical resonances obtained 
by 12th order cycle-expansion using the complete set of 69706 primitive
periodic orbits with symbol lengths $n_p\le 12$.
The exact quantum resonances have been obtained in the region
$0\le {\rm Re}~k\le 60$ by solving Eq.~(\ref{M_4sphere}) with matrix 
dimensions up to $(1751\times 1751)$.

In the region ${\rm Re}~k\lesssim 60$ the resonances obtained by the
two semiclassical methods are in excellent agreement except for the 
imaginary parts of some resonances very deep down in the complex plane.
In this region the semiclassical resonances agree well with the exact
quantum mechanical resonances, the deviations are due to the semiclassical 
approximation, i.e., the first-order $\hbar$ expansion in the semiclassical
trace formula.
As in the case $R=6a$ (Sec.~\ref{d6:sec}) some quantum resonances in the
region ${\rm Re}~k\lesssim 10$ are related to the diffraction of waves at 
the spheres and do not have a semiclassical analogue without the appropriate
extension of the periodic orbit theory \cite{Vat94,Ros96}.
At ${\rm Re}~k\gtrsim 60$ the agreement between resonances obtained
semiclassically via cycle-expansion and harmonic inversion becomes less
perfect, especially for some broad resonances with ${\rm Im}~k\lesssim -1.1$.
Unfortunately, no quantum results are currently available for ${\rm Re}~k > 60$
to judge the quality and accuracy of the semiclassical computations in that
region.

\subsection{Four touching spheres ($R=2a$)}
\label{d2:sec}
The semiclassical quantization of the 4-sphere system becomes even more
difficult when the spheres are further moved together and the symbolic 
dynamics becomes pruned (see Sec.~\ref{pruning:sec}).
In particular, the case of touching spheres with $R=2a$ is a real challenge
for the following reason.
For touching spheres the symbolic dynamics is pruned in a similar way as
in the 3-disk problem \cite{Han93}.
The closed 3-disk billiard is a bound system, and some eigenenergies have
been extracted by either combining the cycle-expansion method with a 
functional equation \cite{Tan91} or by the harmonic inversion method
\cite{Wei02a,Wei02b}.
However, contrary to the closed 3-disk system the four touching spheres 
do not form a bound system, which means that the method of Ref.~\cite{Tan91}
combining the cycle-expansion method with a functional equation cannot be
applied, and thus the touching 4-sphere system cannot be quantized with the
help of the cycle-expansion method.
Nevertheless, we will now demonstrate that the harmonic inversion method 
applied to a cross-correlated periodic orbit signal can reveal at least 
some of the low-lying semiclassical resonances.

For the construction of the periodic orbit signal we have calculated
about 2.8 million orbits of the touching 4-sphere system with lengths
$L < L_{\rm max} = 3.6$.
(Note that the signal is incomplete as discussed in Sec.~\ref{pruning:sec}.)
For the application of the cross-correlation technique we use the operators
$\hat A_1=1$ (the identity), the squared angular momentum $\hat A_2=L^2$,
and the squared distance from the origin, $\hat A_3=r^2$.
Because the signal is incomplete and rather short the results of the
harmonic inversion are less perfectly converged than for the 4-sphere
system with larger sphere separation, i.e., the amplitudes $d_n$ in
Eq.~(\ref{C_qm}) may deviate from the ideal values $d_n=1$ for true physical
resonances and $d_n=0$ for spurious resonances which must be omitted.
As a criterion to accept resonances we have chosen the condition
$|d_n-1| < 0.5$.

\begin{figure}
\begin{center}
\includegraphics[width=0.95\columnwidth]{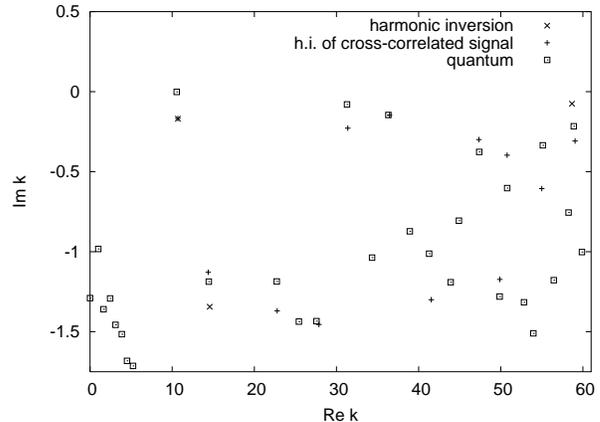}
\end{center}
\caption{$A_1$-resonances in the complex $k$-plane of the touching 4-sphere 
system with radius $a=1$ and center-to-center separation $R=2$. 
Squares: Quantum computations.
Crosses: Semiclassical resonances obtained by harmonic inversion 
without using cross-correlation. 
Plus symbols: Semiclassical resonances obtained by the harmonic inversion of 
a $(3\times 3)$ cross-correlated periodic orbits signal using the operators
1 (identity), $L^2$, and $r^2$.}
\label{fig5}
\end{figure}

The results of our semiclassical and quantum computations for the four 
touching spheres are presented in Fig.~\ref{fig5}.
The crosses mark the semiclassical resonances obtained by harmonic inversion
of the one-dimensional periodic orbit signal.
The low number of crosses indicates that the convergence of the 
one-dimensional signal is not very satisfactory.
The plus symbols show the resonances obtained by harmonic inversion of
the $(3\times 3)$ cross-correlated periodic orbit signal using the operators
$\hat A_1=1$, $\hat A_2=L^2$, and $\hat A_3=r^2$.
With the cross-correlation technique the convergence properties have been
significantly improved compared to the analysis of the one-dimensional signal.
The real parts of the semiclassical resonances agree well with the real
parts of the exact quantum mechanical resonances marked by the squares in
Fig.~\ref{fig5}.
The agreement between the imaginary parts is, however, less perfect.
Some quantum resonances in Fig.~\ref{fig5} do not have a semiclassical
counterpart.
Those resonances with ${\rm Re}~k < 10$ are probably related to the
diffraction of waves at the spheres as discussed above, i.e., they
cannot be explained without extensions of the semiclassical theories
applied in this paper.

\subsection{Efficiency of the semiclassical and quantum algorithms}
\label{efficiency:sec}
For the 4-sphere system as an example of a physical system with three
degrees of freedom we now wish to discuss and compare the efficiency of 
the semiclassical and quantum computations.
As mentioned in the introduction (Sec.~\ref{intro:sec}) the efficiency
of quantum computations usually decreases rapidly with the number of
degrees of freedom of the physical system.
It is an interesting and important question whether semiclassical methods
can beat the efficiency of quantum computations with increasing dimension
of the problem.
Although there is not much hope and evidence that this is generally true,
because of the exponential proliferation of periodic orbits in chaotic systems,
it can be true for certain specific systems.
An example for the superiority of semiclassical over quantum mechanical
calculations is the 4-sphere system with large sphere separation, e.g.\ 
$R=6a$, where semiclassical resonances can easily be obtained even in 
energy regions which are out of reach for the presently known quantum 
techniques \cite{Mai02}.
To understand this it is instructive to study the expense and scaling 
properties of the quantum and classical computations for the 3-disk and 
4-sphere system.

As explained in Sec.~\ref{qm:sec} exact quantum resonances of the 3-disk
and 4-sphere systems can be obtained as roots of Eqs.~(\ref{M_3disk}) and
(\ref{M_4sphere}), respectively, with angular quantum numbers truncated
at $l_{\rm max} \gtrsim 1.5\, ka$.
The calculation of the matrix elements 
${\mathbf M}(k)_{lm,l'm'}^{\textrm{\scriptsize 4-sphere}}$ in (\ref{M_4sphere})
is much more expensive than for the matrix elements
${\mathbf M}(k)_{mm'}^{\textrm{\scriptsize 3-disk}}$ in (\ref{M_3disk}).
However, the serious problem of solving Eq.~(\ref{M_4sphere}) is the 
scaling of the dimension of the matrix 
${\mathbf M}(k)_{lm,l'm'}^{\textrm{\scriptsize 4-sphere}}$, which is
an $N\times N$ matrix with $N=(l_{\rm max}+2)(l_{\rm max}+3)/6$, i.e.,
$N$ scales as $N\sim k^2$ for the 4-sphere system, as compared to
$N\sim k$ for the 3-disk system, Eq.~(\ref{M_3disk}).
For example, in the region $ka\approx 200$ the required matrix dimension is 
$N\gtrsim 300$ for the 3-disk as compared to $N\gtrsim 15000$ for the 
4-sphere system.
With currently available computer technology it is, therefore, impossible
to significantly extend the quantum calculations for the 4-sphere system 
to the region $ka\gg 60$ using Eqs.~(\ref{M_4sphere}-\ref{C_def}).
Note that the cost of the quantum computations does not depend on the
separation $R$ between the disks or spheres.

The expense of the semiclassical quantization is basically given by the
required number of periodic orbits which, in chaotic systems, increases
exponentially with the symbolic or physical length of the orbits.
For the 3-disk system the number of symmetry reduced primitive periodic
orbits with symbol length $n_p$ is given approximately by $N\sim 2^{n_p}/n_p$
whereas it scales as $N\sim 3^{n_p}/n_p$ for the 4-sphere system. 
Contrary to the quantum computations the numerical expense for the 
semiclassical quantization, i.e., the required number of orbits depends 
on the separation $R$ between the disks or spheres.
For large separation $R=6a$ the cycle-expansion method is most efficient 
for the calculation of a large number of resonances.
The reason is that the assumption of the cycle-expansion that the 
contributions of longer periodic orbits in the expansion of the
Gutz\-willer-Voros zeta function (\ref{Z_GV}) are shadowed by pseudo-orbits
composed of shorter periodic orbits is very well fulfilled.
The harmonic inversion method also allows for the calculation of a large
number of resonances, but requires a larger input set of periodic orbits.
While for the two-dimensional 3-disk system the semiclassical and quantum 
computations are very efficient, the semiclassical methods are superior to
the quantum techniques for the three-dimensional 4-sphere system.
The semiclassical calculations can easily be extended to the region
${\rm Re}~ka\gtrsim 60$ where no quantum results are available because 
of the unfavorable scaling of the dimension of the matrix 
${\mathbf M}_{lm,l'm'}$ in Eq.~(\ref{M_4sphere}).
Of course, a more efficient quantum method for the 4-sphere system than 
that of Ref.~\cite{Hen97} may in principle exist.
However, to the best of our knowledge no such method has been proposed 
in the literature to date.
The 4-sphere system therefore is an example of a three-dimensional system
where semiclassical methods are presently superior to exact quantum
calculations.

At reduced separation $R=2.5\, a$ between the disks or spheres the 
semiclassical quantization requires an increased set of periodic orbits to 
achieve convergence of the cycle-expansion or harmonic inversion analysis.
However, for the 4-sphere system the semiclassical methods are still
superior to the exact quantum computations, i.e., semiclassical resonances
can be obtained in regions which are unattainable with the quantum methods
as can be seen in Fig.~\ref{fig4}.

The situation is different for touching spheres, $R=2a$, which is a challenging
system not only for the quantum but also for the semiclassical computations.
The construction of a long periodic orbit signal is impossible because
orbits with increasing sequences of consecutive `0' symbols in the code 
lead to accumulation points in the physical length similar as for the
closed 3-disk system \cite{Wei02a,Wei02b}.
The semiclassical calculations for the touching spheres are therefore at 
least about the same or even more expensive than the quantum computations.

\section{Conclusions}
\label{conclusion:sec}
In summary, we have investigated an open system with three degrees of freedom,
viz.\ the 4-sphere scattering problem with various sphere separations by means
of classical, semiclassical, and quantum mechanical methods.
The classical system has genuinely three-dimensional periodic orbits.
In the symmetry reduced fundamental domain, they can be associated to a ternary
symbolic alphabet, which allows for a systematic periodic orbit search.
For large separations between the spheres ($R\gtrsim 2.5\, a$) semiclassical
resonances have been obtained by application of the cycle-expansion technique
and the harmonic inversion method.
For touching spheres ($R=2a$), the symbolic dynamics is pruned and the 
cycle-expansion does not converge, however, some semiclassical resonances can
be revealed by harmonic inversion of a cross-correlated periodic orbit signal.

Exact quantum mechanical resonances have also been calculated, however,
the quantum computations for the three-dimensional 4-sphere system are
much more expensive than for the two-dimensional analogue, viz.\ the
3-disk scattering problem.
Therefore, the quantum computations had to be restricted to the region
with relatively low wave numbers, i.e., ${\rm Re}\, ka < 60$.
By analyzing the scaling properties of both the quantum and semiclassical
calculations we have demonstrated the superiority of semiclassical methods 
over quantum computations at least for large sphere separations, i.e., 
semiclassical resonances can easily be obtained in energy regions which 
at present are unattainable with the established quantum method.
These results may encourage the investigation of other systems with three or
more degrees of freedom with the goal of developing powerful semiclassical
techniques, which are competitive with or even superior to quantum
computations for a large variety of systems.

In those regions where exact quantum results for the 4-sphere system are
lacking an assessment of the accuracy of the semiclassical resonances is
presently impossible.
Higher-order $\hbar$ corrections have been calculated for two-dimen\-sion\-al 
billiard systems \cite{Gas93,Vat96,Wei02c}, however, the extension of the 
theory to three-dimen\-sion\-al systems is a nontrivial task for future work.

Those quantum resonances which are related to diffraction of waves at the
spheres have not yet been explained semiclassically.
For the 3-disk system diffractive resonances have been obtained with
an extended periodic orbit theory by including the contributions of 
creeping orbits \cite{Vat94,Ros96}.
It will be interesting to generalize these ideas to the genuinely 
three-dimensional 4-sphere system.

\acknowledgments
This work was supported by the National Science Foundation (PHY0071742),
Deutsche Forschungsgemeinschaft (SFB~382),
and the Deut\-scher Akademi\-scher Austauschdienst.
E.A.\ thanks J.M.\ and G.W.\ for the kind hospitality at the Institut f\"ur 
Theoretische Physik during his stay in Stuttgart.


\end{document}